\documentclass[aps,prl,twocolumn,showpacs,superscriptaddress,
preprintnumbers,amsmath,amssymb]{revtex4}
\usepackage{graphicx}% Include figure files
\usepackage{dcolumn}% Align table columns on decimal point
\usepackage{bm}% bold math
\usepackage{ifthen}
\usepackage{booktabs}

\begin{document}

\title{Atomistic theory of spin relaxation in self-assembled 
In$_{1-x}$Ga$_x$As/GaAs quantum dots at zero magnetic field}

\author{Hai Wei}
\affiliation{Key Laboratory of Quantum Information, University of
Science and Technology of China, Hefei, 230026, People's Republic of
China}

\author{Ming Gong}
\affiliation{Key Laboratory of Quantum Information, University of
Science and Technology of China, Hefei, 230026, People's Republic of
China}

\author{Guang-Can Guo}
\affiliation{Key Laboratory of Quantum Information, University of
Science and Technology of China, Hefei, 230026, People's Republic of
China}

\author{Lixin He \footnote{Email address: helx@ustc.edu.cn}}
\affiliation{Key Laboratory of Quantum Information, University of
Science and Technology of China, Hefei, 230026, People's Republic of
China}

\date{\today }

\begin{abstract}

We present full atomistic calculations of the spin-flip time (T$_{1}$) of electrons and holes
mediated by acoustic phonons in self-assembled In$_{1-x}$Ga$_x$As/GaAs quantum dots
at zero magnetic field. At low magnetic
field, the first-order process is suppressed, and the second-order process
becomes dominant. We find that the spin-phonon-interaction induced spin
relaxation time is 40 - 80 s for electrons, and 1 - 20 ms for holes at 4.2 K. The
calculated hole-spin relaxation times are in good agreement with recent experiments, 
which suggests that the two-phonon process is the main relaxation mechanism for hole-spin
relaxation in the self-assembled quantum dots at zero field. We further clarify the
structural and alloy composition effects on the spin relaxation in the quantum dots.
\end{abstract}



\maketitle

The electron/hole spin in semiconductor quantum dots (QDs) is 
believed to be a promising candidate for solid state quantum 
computations \cite{loss98}. Recently years, huge progress 
have been made experimentally 
in initialization, manipulation and controlling of the 
spins in QDs\cite{kroutvar04,braun05,heiss07,gerardot08}. 
However, the short spin lifetime 
is still a major obstacle to realize the quantum computation. Until now the
mechanisms for spin relaxation in QDs
are still not well understood both in theory and in experiments. 
Spin-phonon interaction due to the spin-orbit coupling (SOC) effects
is one of the main
mechanisms lead to spin relaxation in the QDs \cite{khaetskii01,golovach04}, 
which depends strongly on the geometries and compositions
of the QDs. An accurate description of SOC is 
crucial to understand the spin relaxation. Unfortunately, the understanding of the spin
relaxation from the atomistic level is still unavailable.

Electrons were expected to have very 
long spin lifetime, because they have small
SOC in QDs. However, because of hyperfine interactions with nuclear
spins, the electron spin coherence time is greatly reduced ($ \sim $ 500 ps) \cite{braun05}. 
The electron spin lifetime can be prolonged by applying an external
magnetic field to polarize the nuclear spin \cite{braun05,kroutvar04}. 
Unfortunately, at the same time, 
the electron spin lifetimes (T$_{1}$) decrease fast with the magnetic field,
being proportional to B$^{-5}$ \cite{kroutvar04}.
In contrast to electrons, holes have $p$-like atomic 
wave functions. Therefore, the hyperfine interactions with the nuclear spins are
small \cite{eble09}. 
Indeed, recent experiments show that holes have very long spin
lifetimes in InAs/GaAs QDs. 
For example, Heiss et al. found that the holes can have lifetimes of
0.27 ms at 1.5 T \cite{heiss07}. 
Only very recently, high fidelity spin states have been prepared at low magnetic
field, taking the advantages of the extreme long hole spin lifetimes ($\sim$ 1 ms) \cite{gerardot08}. 

At very low magnetic field, the first-order spin-phonon interaction is greatly
suppressed, and the multi-phonon process becomes dominate. 
The $T_1 \propto B^{-5}$ law breaks down at low magnetic field \cite{kroutvar04,trif09}.
Most previous theoretical works focus on spin relaxation
in a large magnetic field \cite{khaetskii01,golovach04,cheng04}, 
and there are only few studies of spin relaxation in QDs at low magnetic field \cite{trif09}.
These studies are all based on effective mass approximations or $\mathbf{k} \cdot \mathbf{p}$
theory \cite{khaetskii01,cheng04,bulaev05}, 
in which the {\it effective} SOC are added in by hand in the form of 
Dresselhaus interactions and Rashba interactions.
These continuum theories treat poorly the local strain and alloy composition
effects especially for hole, which may play 
an extremely important role to the effective
SOC.

In this letter, we present the first atomistic calculations \cite{williamson00} of 
the spin relaxation of electrons and holes at zero magnetic field in self-assembled In$_{1-x}$Ga$_x$As/GaAs QDs. 
In this method, the Hamiltonian naturally takes the {\it full} SOC
effects into account at the atomistic level. 
The method needs no fitting parameters to the
QD experiments, and therefore can give quantitative predictions
that can be directly compared to experiments.
Our calculations show that the holes have extremely long spin lifetimes, 
1--20 ms at $T$=4.2 K, which is in excellent agreement with the recent experimental
data \cite{gerardot08}. This confirms that the two-phonon scattering process with the SOC
is the main mechanism for the spin relaxation of holes at zero magnetic field.

At low magnetic field, the first-order process is suppressed, 
because the Zeeman splitting between spin-up
and spin-down states is very small. 
Therefore, the density of phonon states that satisfy energy conservation
is extremely low. In this case, the spin relaxation via the two-phonon
processes become dominant:
a spin at the initial (labeled as $i$) state, $\epsilon_i$, absorbs a phonon of
momentum ${\bf q}$ and jumps to an intermediate ($s$) state,
$\epsilon_s$. It emits a phonon with momentum ${\bf k}$ 
and relaxes to the final ($f$) state, $\epsilon_f$, 
which has an opposite spin of the initial state.
The electron and hole spin-flip
rate ($ \tau_{\nu}^{-1} $) from the initial state to the final
state is given by the second-order Fermi's Golden Rule, 
\begin{eqnarray} 
{1 \over \tau_{\nu}}
&=&
\frac{2\pi}{\hbar}\sum_{\mathbf{q,k}}%\sum_{\mathbf{k}}
\left[ \sum_{s}'
(\frac{M_{\mathbf{q}}^{is}M_{\mathbf{k}}^{sf}}{\epsilon_{i}-\epsilon_{s}+\hbar \omega_{\mathbf{q}}} 
+\frac{M_{\mathbf{k}}^{is}M_{\mathbf{q}}^{sf}}{\epsilon_{i}-\epsilon_{s} - \hbar \omega_{\mathbf{k}}})\right]^{2} \nonumber \\
&& N_q (N_k+1) 
\delta(\epsilon_f-\epsilon_i-\hbar \omega_{\mathbf{q}} + \hbar \omega_{\mathbf{k}}) \, .
\label{eq:lifetime}
\end{eqnarray}
where only long-wave acoustic phonons are involved in the process,
$\omega_{\mathbf{q}} = c_\nu|\mathbf{q}|$, and $c_\nu$
are the sound speeds for the $\nu$= LA and TA modes.
$N_q =1/[\exp(\hbar \omega_{\mathbf{q}}/k_{B}T)-1]$ is the
Bose-Einstein distribution function. The summation is over the
intermediate states includes all the states except for the initial
and final states. 
The matrix elements $M_{\mathbf{q}}^{if}$ are given by:
\begin{equation}\label{Mq}
M_{\mathbf{q}}^{if}= \alpha_{\nu}({\bf
  q})\langle\psi_{f}|e^{-i\mathbf{q}\cdot\mathbf{r}}|\psi_{i}\rangle \, ,
\end{equation}
where $|\psi_{i}\rangle$ and $|\psi_{f}\rangle$ are the wave functions of initial
and final state, respectively.
$ \alpha(\mathbf{q}) $ is the electron-phonon coupling strength. 
We have considered three electron-phonon interaction mechanisms 
in the QDs \cite{cheng04}: 
(i) electron-acoustic-phonon interaction due to the deformation
potential ($\nu$ = LADP),
(ii) electron-acoustic-phonon interaction due to the piezoelectric 
field for the longitudinal modes ($\nu$ = LAPZ), 
and (iii) electron-acoustic-phonon interaction due to the piezoelectric 
field for the transverse modes ($ \nu $ = TAPZ).
The parameters are taken from Ref. \cite{cheng04}. 
The overall spin relaxation time $T_1=\sum_{\nu} 1/\tau_{\nu}$.

To calculate $T_1$, it is crucial to obtain high-quality single-particle
wave functions and energy levels \cite{cheng04}. In this work, we use an
atomistic pseudopotential method, which has been proven to be very accurate
for InAs/GaAs QDs \cite{williamson00, he05d, he06a}.
We consider realistic lens-shaped  In$_{1-x}$Ga$_x$As/GaAs QDs embedded in a GaAs matrix containing
60$\times$60$\times$60 8-atom unit cells. 
The dots are assumed to grow along the [001] direction, on
the top of the one-monolayer wetting layers. 

We first relax the dot+box system by minimizing the strain energy as a function
of the coordinate $\{ {\bf R}_{n,\alpha} \}$ of the $\alpha$-th atom at the site
$n$ for all atoms, using the valence force field (VFF) method\cite{keating66,martin70}.
Once we have all the atom positions, we obtain the energy levels and wavefunctions by
solving the single-particle Schr\"{o}dinger equation \cite{footnote1},
\begin{equation}
\left[ -{\nabla^2 \over 2}  + \sum_{n, \alpha} \hat{v}_{\alpha} ({\bf r} - {\bf
  R}_{n,\alpha}) \right] \psi_i ({\bf r})   = \epsilon_i \psi_i ({\bf r}) \, .
\end{equation}
$\hat{v}_{\alpha}({\bf r})$ is the screened atomic pseudopotential for the
$\alpha$-th element, 
including a local potential and a non-local SOC term \cite{williamson00},
\begin{equation}
V_{\rm SO,\alpha}= \sum_l | l \rangle \delta V_{l,\alpha} ({\bf r}) {\bf L} \cdot {\bf S} \langle l|
\, ,
\end{equation}
where $l$ is the angular momentum (only $l$=1 is used).
${\bf L}$ is the angular momentum operator, and ${\bf S}$ is the  
spin operator. The atomic pseudopotentials are fitted to the bulk properties. Once the
parameters are determined, there are no free parameters for modeling the
QDs. The method naturally includes the Rashba and Dresselhaus SOC
in a ``first-principles'' manner.

The single-particle Schr\"{o}dinger equation is solved by
the linear combination of bulk bands(LCBB) method \cite{wang99b}.
We use eight bands for both the electrons and holes, and take the inter-band
coupling (heavy-hole-light-hole coupling and the valence-conduction band
coupling) into account, which is very important for hole/electron spin relaxation.
A 6$\times$6$\times$16 k-mesh converges very well the results. 
Due to the SOC, the wave functions mix opposite spin components,
i.e., $\psi_i = \alpha |\uparrow \rangle + \beta |\downarrow \rangle$.  

To calculate the matrix elements, $M_{\bf q}$, we first calculate
the single-particle wave functions on a sparse $k$ mesh following
Ref. \onlinecite{bester07}. We then 
interpolate the wave functions into a more dense 49$\times$49$\times$129 
$k$-mesh to get accurate results. We sum over 20 intermediate states ($s$ in 
Eq. \ref{eq:lifetime}), which converge a spin lifetime $<$ 1 s for electrons
and $<$ 1 ms for holes. Details on how to calculate $M_{\bf q}$ will be
published elsewhere.

\begin{table}
\centering
\caption{Spin relaxation times of the electrons ($T_{1}^{\rm e}$) and the holes
($T_{1}^{\rm h}$) in self-assembled In$_{1-x}$Ga$_x$As/GaAs QDs 
at 4.2 K. All lengths are in nm. $\beta^{2}$ is the weight of
the minority spins of the initial states.}
\begin{tabular}{ccccccc}
\hline
\hline
     height & base & $x$  & $T_{1}^{\rm e}$ (s) & $\beta^{2}$ ($\times10^{-4}$) &
     $T_{1}^{\rm h}$ (ms) & $\beta^{2}$ ($\times10^{-4}$) \\
        \hline
    2.5 & 20 & 0       & 81 & 19   & 17.9  & 96 \\
    3.5 & 20 & 0       & 59 & 16   & 4.5    & 140 \\
    2.5 & 25 & 0       & 54 & 17   & 20.4  & 65 \\
    3.5 & 25 & 0       & 43 & 16   & 12.4  & 83 \\
    2.5 & 25 & 0.15 & 44 & 19   & 15.5  & 62 \\
    3.5 & 25 & 0.15 & 40 & 11   & 6.4    & 83 \\
    2.5 & 25 & 0.30 & 44 & 20 & 10.8 & 65 \\
    3.5 & 25 & 0.30 & 40 & 10   & 4.5    & 116 \\
    2.5 & 25 & 0.50 & 53 & 3      & 5.2    & 87 \\
    3.5 & 25 & 0.50 & 43 & 5      & 2.7    & 88 \\
\hline % inserts single-line
\hline % inserts single-line
\label{tab_lifetime}
\end{tabular}
\end{table}

In Table \ref{tab_lifetime}, we list the electron and hole spin lifetimes 
for some typical In$_{1-x}$Ga$_x$As/GaAs QDs at $T$=4.2 K. 
We also list the weight of the minority spin of the initial states $\beta^2$ in the Table.
From the Table, it can be seen that the electron spin lifetimes due to the spin-phonon interaction
are extremely long, 40--80 s, because the spin mixture 
is quite small (See Table \ref{tab_lifetime}). 
We find that the inter-band coupling has a large contribution to the spin relaxation for
electrons. Ignoring inter-band coupling leads to a spin lifetime that is about three times longer 
for electrons. However, the actual electron-spin lifetimes are determined by the
hyperfine interaction with nuclear spins at low magnetic field.

The hole-spin lifetimes of typical In$_{1-x}$Ga$_x$As/GaAs QDs are 1 -- 20 ms.  
For alloy dots, the typical hole-spin lifetime is a few ms,
which is in good agreement with recent experimental data ($\sim$ 1 ms) \cite{gerardot08}. 
This suggests that two-phonon spin scattering is the main mechanism that leads
to spin flip for holes in these self-assembled QDs at low magnetic field. 
This finding differs greatly from the results given in Ref. \cite{bulaev05} where the
second-order phonon process is not considered.
The spin lifetimes of the holes are about three orders of magnitudes shorter 
than those of the electrons. This is because
the single-particle energy spacing of the electrons is 3--5 times larger
than that of holes, and the spin mixing due to the
SOC for holes is also much stronger
than that of electrons, as shown in Table \ref{tab_lifetime}. 
The spin-flip times in pure QDs are significantly longer
than those of alloy QDs. The main reason for the longer hole spin-flip time in pure
QDs is that the energy spacing in pure QDs is larger, and the spin mixture in alloy
dots tends to be slightly larger than that in the pure dots.  
 
\begin{figure}
\centering
\includegraphics[width=2.8 in]{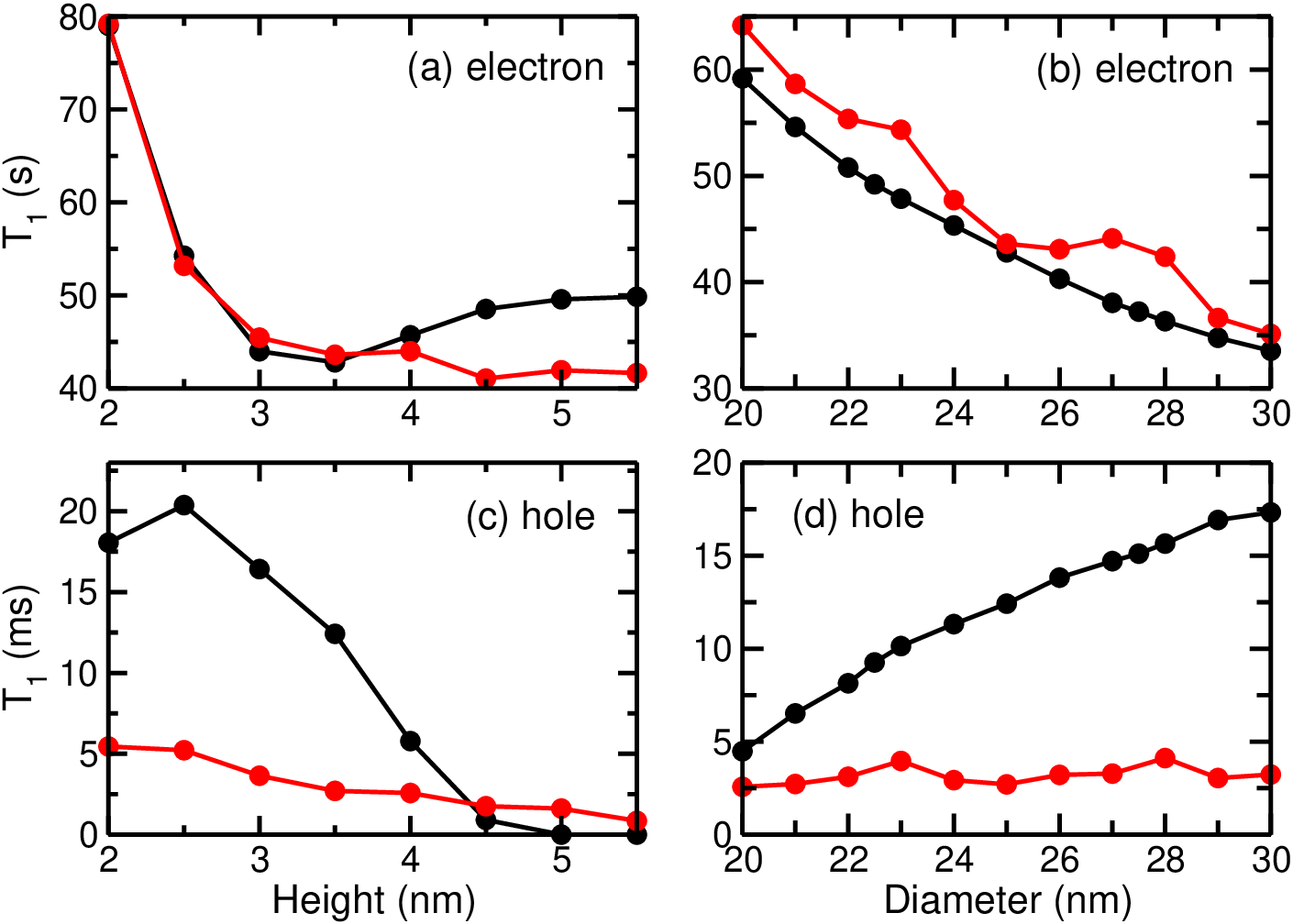}
\caption{(Color online) Spin relaxation times of electrons and holes in lens-shaped
  In$_{1-x}$Ga$_x$As/GaAs QDs of different heights [(a), (c)], and
  base diameters [(b), (d)]. Height varies keeping diameter at 25 nm and
  the diameter varies keeping height at 3.5 nm. Black lines are the results for
pure dots, whereas the red lines are the results for ally dots with $x$=0.5.}
\label{fig:geometry}
\end{figure}

The atomistic theory reveals some important features 
that can not be observed by $k\cdot p$ method, such as geometry and
alloy compositions effects to the spin relaxation.
To explore the effects of geometry on the spin flip time in QDs, we calculate the
spin flip time of lens-shaped QDs, as functions of dot height (diameter) while
fixing the dot diameter (height). 
The results are presented in Fig. \ref{fig:geometry} (a) and (b)
and for electrons, and Fig. \ref{fig:geometry} (c) and (d) for holes.
For pure dots (black dots), the spin lifetime of the electrons decreases as dot height increases 
from 2 nm to 3.5 nm, and becomes flat for taller dots.
The spin lifetimes of holes, also decreases with increases in the height of the dots. For
extremely tall pure InAs/GaAs quantum dots, the spin flip time is extremely fast
(about two orders of magnitudes faster). This is because in very tall pure
InAs/GaAs QDs, holes tend to localize in the interface between the dots and
matrix, resulting in a very small energy spacing, and a fast spin relaxation. 
The spin lifetimes decrease with increasing dot diameter for electrons
[Fig. \ref{fig:geometry} (c)]. In contrast, the spin lifetimes increase with
increasing dot diameter for holes [Fig. \ref{fig:geometry} (d)].
To understand the effects of geometry on the spin lifetime, we note that there
are two competing factors: the energy spacing and spin mixing. For electrons, the 
SOC is not very sensitive to the dot geometry, and the spin
lifetimes are mainly determined by the single-particle energy spacing. 
For holes, increasing dot height leads to a smaller energy
spacing between the levels, and the spin mixing also increases with dot height because
of the stronger effective SOC. Therefore the spin lifetime decreases.
Increasing the dot diameter, however, leads to a smaller energy
spacing between the levels, which tends to increase the spin flip time,
but the spin mixing decreases with increasing dot diameter. The overall
effect is that the spin lifetime increases with increasing dot diameter.
Therefore, for pure InAs/GaAs QDs, a flatter QD has a much longer lifetime than a tall QD.

We also show that the geometry-dependent spin lifetime for
In$_{0.5}$Ga$_{0.5}$As/GaAs alloy
QDs (red dots) in Fig. \ref{fig:geometry}. 
For electrons, the spin lifetimes of alloy QDs are very similar to those of
pure QDs, whereas for holes, 
the alloy QDs show a much weaker geometry dependence for their spin
lifetime. When the dot height increases, the lifetime decreases slightly with
increasing dot height. However, for most of the dots, the alloy dots
have shorter lifetimes than those of the pure dots. For very tall dots, the
alloy dots have longer lifetime, because the holes are not localized at the
dot-matrix interface in the alloy dots. 
The spin lifetimes of holes are also flat with increasing of the dot diameter.

\begin{figure}
\centering
\includegraphics[width=2.8 in]{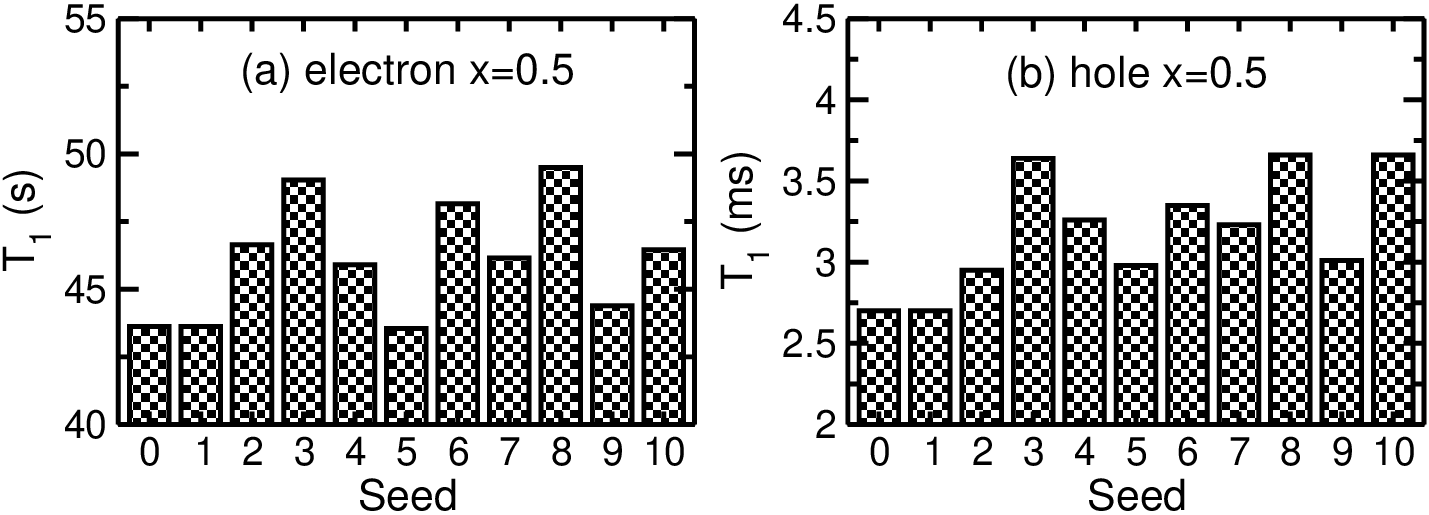}
\caption{Spin relaxation time of In$_{0.5}$Ga$_{0.5}$As/GaAs QDs with
  different Ga distributions.}
\label{fig:alloy}
\end{figure}

It has been shown by Mlinar and Zunger \cite{mlinar09}
that the randomness of In, Ga alloys may have large effects on their optical
properties for finite sizes In$_{1-x}$Ga$_x$As/GaAs QD ($<$ 10$^5$ atoms). 
To investigate the alloy randomness effects on the spin relaxation time, 
we calculate $T_1$ for In$_{0.5}$Ga$_{0.5}$As/GaAs QDs, 
with various random distributions of Ga atoms in the dots. 
The results are
shown in Fig.~\ref{fig:alloy}.
We find that the spin lifetimes of electrons are distributed between 44 s to 50
s, in which the lifetime fluctuation is less than 10 \%. 
The spin lifetimes of the holes are distributed between 2.7 ms to 3.6 ms and the 
fluctuation of the lifetimes of holes is about 33 \%.
Analyses show that the spin lifetime fluctuation mainly comes from fluctuations in
the single-particle energy spacing (0.4 meV
for electrons and 1.0 meV for holes) caused by the Ga distributions.

\begin{figure}
\centering
\includegraphics[width=2.8 in]{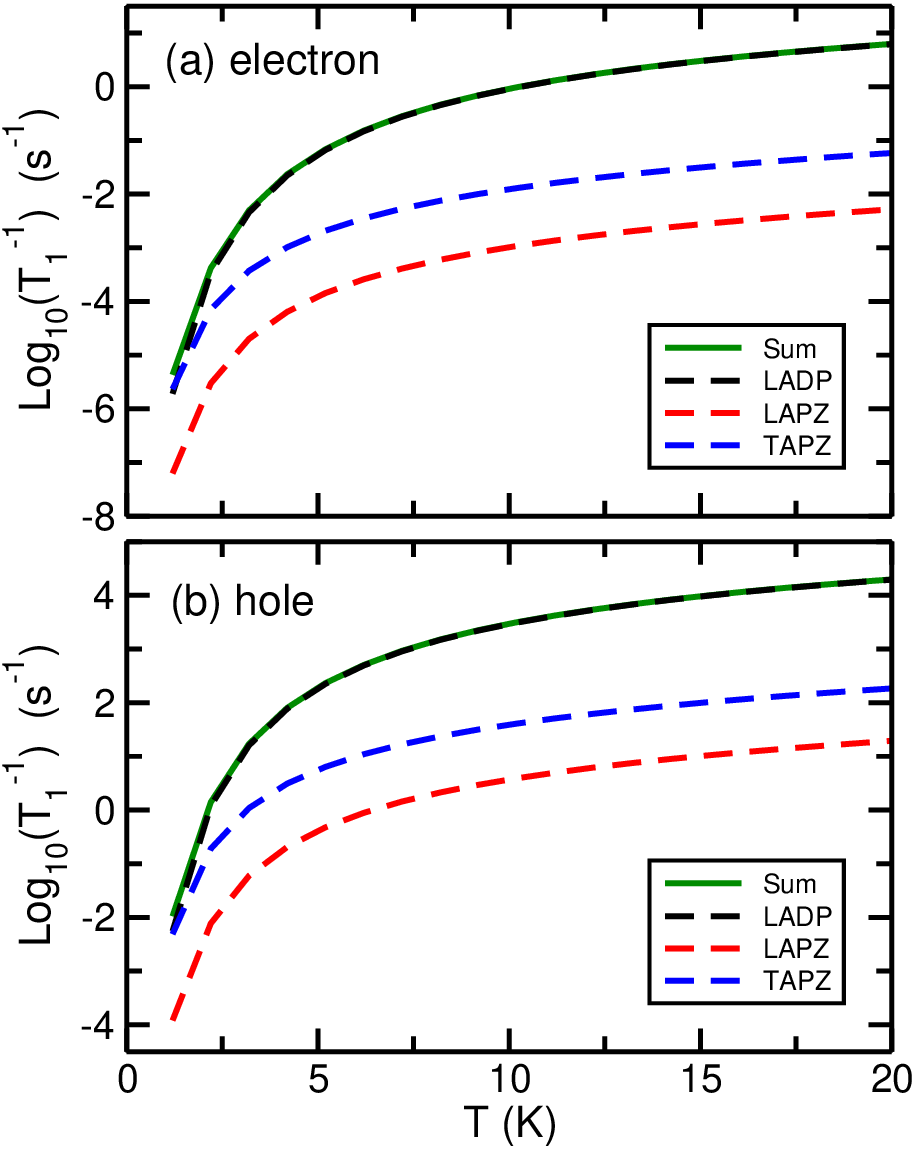}
\caption{(Color online) The electron- (a) and hole- (b) spin relaxation rates
  (1/$T_{1}$) of a pure InAs/GaAs QD as functions of temperature. QD base diameter=25 nm, height=3.5 nm.
  The black, red, and blue lines represent the spin relaxation
  rate due to the LADP, LAPZ, and TAPZ mechanisms, respectively. The green solid lines are the total
  relaxation rates.}
\label{fig:temperature}
\end{figure}

Figure \ref{fig:temperature} depicts the spin-flip rates (green solid lines) 
for electrons and holes as functions temperature for a lens-shaped InAs/GaAs QD
with diameter $D$=25 nm, and height, $h$=3.5 nm. 
The spin relaxation times for both electrons and holes
increase quickly with decreasing temperature at low
temperatures \cite{trif09}, because fewer phonons can take part in the scattering process.
At lower temperatures, the holes may have extremely long lifetimes, for example, 
$T$=2 K, $T_1$=1.5 s, suggest that hole spins can be promising candidates for
quantum computing. 
The spin relaxation times $ T_1 \sim 1/T^{2} $ for temperatures $ > $20 K, which is
different from the one phonon process, where $T_1  \sim 1/T$ \cite{trif09}.
We also compare the contributions of the three carrier-phonon interactions to the
spin flip time. Among the three mechanisms, the contributions from the
LAPZ (red dashed lines) and the TAPZ (blue dashes lines) are 2 to 3 orders of
magnitudes weaker than that of the LADP (black dashed lines) 
from 1K to room temperature (not shown).
Below 1 K, the TAPZ has similar contribution to the spin flip rate to LADP.

%\section{Summary}

To summarize, we presented the first atomistic calculations of 
the multi-phonon-induced electron/hole 
spin relaxation in self-assembled  In$_{1-x}$Ga$_x$As/GaAs
quantum dots at zero magnetic field.  
The calculated the spin lifetimes of the holes were in the range of a few milliseconds to
20 milliseconds at 4.2 K, which is in excellent agreement with recent experimental results.
We further studied the structural and alloy composition effects on the spin
lifetimes, and found that the alloy QDs have much shorter hole-spin lifetimes.
Compared with previous methods, the current method has greater predictive power,
and we expected it to play an important role in future studies on spin
relaxations in quantum dots.

LH acknowledges support from the Chinese National
Fundamental Research Program 2011CB921200, ``Hundreds of Talents'' program 
from Chinese Academy of Sciences 
and National Natural Science Funds for Distinguished Young Scholars.

%\bibliographystyle{apsrev}
%\bibliography{MyDotBiblio,DotBiblio,footnote} 

\begin{thebibliography}{20}
\bibitem[{foo()}]{footnote1}
\bibinfo{note}{In the calculations, we apply a $B=10 ^{-3}T$ magnetic field
  along growth direction to separate the spin-up and spin-down states. For such
  small magnetic field, we consider only the Zemann splitting due to the
  magnetic field, and neglect the gauge terms.}
\end{thebibliography}

\end{document}